\input harvmac
%\draftmode
\noblackbox

\input epsf

\def\tilde{\widetilde}
\def\hat{\widehat}
\newcount\figno
\figno=0
\def\fig#1#2#3{
\par\begingroup\parindent=0pt\leftskip=1cm\rightskip=1cm\parindent=0pt
\baselineskip=11pt
\global\advance\figno by 1
\midinsert
\epsfxsize=#3
\centerline{\epsfbox{#2}}
\vskip 12pt
{\bf Fig.\ \the\figno: } #1\par
\endinsert\endgroup\par
}
\def\figlabel#1{\xdef#1{\the\figno}}
\def\encadremath#1{\vbox{\hrule\hbox{\vrule\kern8pt\vbox{\kern8pt
\hbox{$\displaystyle #1$}\kern8pt}
\kern8pt\vrule}\hrule}}

\def\half{{1\over 2}}

\def\b{{\beta}}

\def\a{{\alpha}}

\def\D{{\Delta}}
\def\m{{\mu}}
\def\n{{\nu}}
\def\ep{{\epsilon}}
\def\d{{\delta}}

\def\ph{{\phi}}
\def\t{{\theta}}
\def\l{{\lambda}}

\def\Ph{{\Phi}}

\def\vol{{\rm vol}}

\def\r{{\rightarrow}}

\def\frac#1#2{{#1\over #2}}

\def\L{{\Lambda}}
\def\T{{\Theta}}

\def\p{\partial}

\lref\rsw{N. Seiberg and E.Witten, ``String Theory and
Noncommutative Geometry,'' hep-th/9908142, JHEP { \bf 09} (1999)
032.} 
\lref\rcds{A. Connes, M. Douglas and A. Schwarz,
``Noncommutative Geometry and Matrix Theory: Compactification on
Tori,'' hep-th/9711162, JHEP { \bf 02} (1998) 003.}
\lref\der{G. Derrick, ``Comments on Nonlinear 
wave equations as models for Elementary Particles,'' 
J. Math. Phys {\bf 5} 1252 (1964). 
S. Coleman, ``Aspects of
Symmetry'', Chapters 6 and 7 } 
\lref\ns{N. Nekrasov and A. Schwarz,
``Instantons on noncommutative $R^4$, and (2,0) superconformal 
six dimensional theory,''
hep-th/9802068, Commun. Math. Phys. {\bf 198} (1998) 689-703.}  
\lref\ikkt{N. Ishibashi, H. Kawai, Y. Kitazawa and  A. Tsuchiya, 
``A Large-N Reduced Model as Superstring,' hep-th/9612115, 
Nucl.Phys. {\bf B498} (1997) 467-491.}
\lref\wsft{E. Witten, ``Noncommutative Geometry and String Field Theory,''
Nucl. Phys. {\bf B268} (1986) 253.} 
\lref\hlrs{G. Horowitz, J. Lykken R. Rohm 
and A . Strominger, ``A purely cubic action for String Field Theory, ''
Phys.Rev.Lett. { \bf 57} (1986) 283-286.}
\lref\bfss{T. Banks, W. Fischler, S.H. Shenker and  L. Susskind, 
``M Theory As A Matrix Model: A Conjecture,''
hep-th/9610043, Phys.Rev. { \bf D55} (1997) 5112-5128.}
%%%%%%%%%%%%%%%%%%%%%%%%%%%%%%%%%%%%%%%%%%%%%%%%%%%

\Title
{\vbox{\baselineskip12pt
\hbox{hep-th/0003160}}}
{\vbox{\centerline{Noncommutative Solitons}}}

\centerline{Rajesh Gopakumar, Shiraz Minwalla and Andrew Strominger}

\centerline{\sl Jefferson Physical Laboratory, Harvard University}
\centerline{\sl Cambridge, MA 02138, USA}
\medskip

\vskip 0.8cm

\centerline{\bf Abstract}
\medskip
\noindent

We find classically stable solitons (instantons) in odd (even)
dimensional scalar noncommutative field theories whose scalar
potential, $V(\ph)$, has at least two minima. These
solutions are bubbles of the false vacuum whose size 
is set by the scale of noncommutativity. Our construction uses the 
correspondence between non-commutative fields and operators on a single
particle Hilbert space.
In the case of noncommutative gauge
theories we note that expanding around a simple solution shifts 
away the kinetic term and results in a purely quartic action with linearly 
realised gauge symmetries.

\vskip 0.5cm
\Date{Mar. 2000}
\listtoc
\writetoc

\newsec{Introduction}

Quantum field theory on a noncommutative space is of interest for
a variety of reasons. It appears to be a
self-consistent deformation of the highly constrained structure of
local quantum field theory. Noncommutative field theories are nonlocal;
unraveling the consequences of the breakdown of locality at short
distances may help understanding non-locality in 
quantum gravity. The discovery of noncommutative quantum field
theory in a limit of string theory \rcds\ provides new inroads to
the subject.

\nref\rfilk{T.~Filk, ``Divergences in a Field Theory on Quantum
Space,'' Phys. Lett. {\bf B376} (1996) 53.}%
\nref\onevgb{J.C.~Varilly and J.M.~Gracia-Bondia, ``On the ultraviolet
behavior of quantum fields over noncommutative manifolds,''
Int.\ J.\ Mod.\ Phys.\ {\bf A14} (1999) 1305, hep-th/9804001.}%
\nref\two{M.~Chaichian, A.~Demichev and P.~Presnajder, ``Quantum Field
Theory on Noncommutative Space-times and the Persistence of
Ultraviolet Divergences,'' hep-th/9812180;  ``Quantum Field Theory on
the Noncommutative Plane with E(q)(2) Symmetry,'' hep-th/9904132.}%
\nref\rjabbari{ M. Sheikh-Jabbari, ``One Loop Renormalizability
of Supersymmetric Yang-Mills Theories on Noncommutative Torus,''
hep-th/9903107, JHEP {\bf 06} (1999) 015.}%
\nref\rruiz{C.P.Martin, D. Sanchez-Ruiz,
``The One-loop UV Divergent Structure of U(1) Yang-Mills
 Theory on Noncommutative $R^4$,''hep-th/9903077,
Phys.Rev.Lett. {\bf 83} (1999) 476-479.}%
\nref\rwulkenhaar{T. Krajewski, R. Wulkenhaar, ``Perturbative quantum
gauge fields on the noncommutative torus,'' hep-th/9903187.}%
\nref\twopfo{S.~Cho, R.~Hinterding, J.~Madore and H.~Steinacker,
``Finite Field Theory on Noncommutative Geometries,''
hep-th/9903239.}%
\nref\three{E.~Hawkins, ``Noncommutative Regularization for the
Practical Man,'' hep-th/9908052.}%
\nref\rsusskind{D. Bigatti and L. Susskind, 
``Magnetic fields, branes and noncommutative geometry,'' 
hep-th/9908056.}%
\nref\aoki{H. Aoki, N. Ishibashi, S. Iso, H. Kawai, Y. Kitizawa
and T. Tada, hep-th/9908141.}
\nref\rishibashi{N. Ishibashi, S. Iso, H. Kawai and Y. Kitazawa,
``Wilson Loops in Noncommutative Yang-Mills,'' hep-th/9910004.}%
\nref\riouri{I. Chepelev and  R. Roiban, 
``Renormalization of Quantum Field Theories on Noncommutative $R^d$, I. 
Scalars,'' hep-th/9911098.}%
\nref\rbenaoum{H. Benaoum, ``Perturbative BF-Yang-Mills theory 
on noncommutative $R^4$,'' hep-th/9912036.}%
\nref\msv{S. Minwalla, M. Van Raamsdonk, N. Seiberg ``Noncommutative 
Perturbative Dynamics,'' JHEP {\bf 02} (2000) 020, hep-th/9912072.}%
\nref\arcioni{G.~Arcioni and M.~A.~Vazquez-Mozo,
``Thermal effects in perturbative noncommutative gauge theories,''
JHEP {\bf 0001} (2000) 028, hep-th/9912140.}%
\nref\haya{M.~Hayakawa,
``Perturbative analysis on infrared aspects of noncommutative QED on
R**4,'' hep-th/9912094;
``Perturbative analysis on infrared and ultraviolet aspects of  
noncommutative QED on R**4,'' hep-th/9912167.}%
\nref\ikk{S.~Iso, H.~Kawai and Y.~Kitazawa,
``Bi-local fields in noncommutative field theory,'' hep-th/0001027.}%
\nref\grosse{H.~Grosse, T.~Krajewski and R.~Wulkenhaar,
``Renormalization of noncommutative Yang-Mills theories: A simple  
example,''
hep-th/0001182.}%
\nref\arefeva{I.~Y.~Aref'eva, D.~M.~Belov and A.~S.~Koshelev,
``A note on UV/IR for noncommutative complex scalar field,''
hep-th/0001215.}%
\nref\texa{W.~Fischler, E.~Gorbatov, A.~Kashani-Poor, S.~Paban,
P.~Pouliot and J.~Gomis, 
``Evidence for winding states in noncommutative quantum field theory,''
hep-th/0002067.}%
\nref\sunew{A.~Matusis, L.~Susskind and N.~Toumbas,
``The IR/UV connection in the non-commutative gauge theories,''
hep-th/0002075.}%
\nref\sv{M. Van Raamsdonk and  N. Seiberg, 
``Comments on Noncommutative Perturbative Dynamics, ''
hep-th/0002186.}
\nref\as{F. Ardalan and  N. Sadooghi, 
``Axial Anomaly in Noncommutative QED on $R^4$,''hep-th/0002143.}%
\nref\chu{C. Chu, ``Induced Chern Simons and WZW Action in Noncommutative 
Spacetime,'' hep-th/0003007.}%
\nref\da{O. Andreev and H. Dorn, ``Diagrams of Noncommutative Phi-Three 
Theory from String Theory,''hep-th/0003113.}%
\nref\sang{Y. Kiem and  S. Lee, ``UV/IR Mixing in Noncommutative 
Field Theory via Open String Loops,''hep-th/0003145.}%
\lref\moyalb{A. Dhar, G. Mandal, and S. R. Wadia, ``Nonrelativistic
Fermions, Coadjoint Orbits of $W_\infty$, and String Field Theory
At $c=1$,'' hep-th/9207011, ``$W_\infty$ Coherent States And
Path-Integral Derivation Of Bosonization Of Non-Relativistic Fermions
In One Dimension,'' hep-th/9309028, ``String Field Theory Of Two
Dimensional QCD: A Realization Of $W_\infty$ Algebra,''
hep-th/9403050.}%
\lref\moyalc{D. B. Fairlie, T. Curtright, and C. K. Zachos, ``Integrable
Symplectic Trilinear Interactions For Matrix Membranes,'' Phys. Lett.
{\bf B405} (1997) 37, hep-th/9704037; D. Fairlie, ``Moyal Brackets In
$M$-Theory,'' Mod. Phys. Lett. {\bf A13} (1998) 263, hep-th/9707190;
``Matrix Membranes And Integrability,'' hep-th/9709042.}%

Perturbative aspects of noncommutative field theories have been
analyzed in \refs{\rfilk - \sang}. This study has thrown up some evidence 
for the renormalizability of a class of noncommutative field 
theories, and has revealed an intriguing mixing of the UV and IR
\msv\ in these theories. 
In this paper we will construct localized 
classical solutions in some simple noncommutative
field theories. We expect these objects 
to play a role in the quantum dynamics of
the theory.

We first consider a scalar field with a polynomial
potential. A scaling argument due to Derrick \der\ shows
that, in the commutative case, solitonic solutions do not exist in
more than $1+1$ dimensions, as the energy of any 
field configuration can always be lowered by shrinking. 
Perhaps surprisingly, for sufficiently large 
noncommutativity parameter $\t$, we will find 
classically stable solitons in any theory with a scalar potential with
more than one $\it local$ minimum. These solitons are asymptotic
to the true vacuum, and reach a second (possibly false) vacuum in
their core. They cannot decay simply by shrinking to zero size
because sharply peaked field
configurations have high energies in noncommutative field theories. 
These solitons are  metastable in the 
quantum theory, but by adjusting parameters in the scalar potential, 
their lifetime can be made arbitrarily long while their mass is kept 
fixed. 
Solutions are found corresponding to solitons in $2l+1$
dimensions or instantons in $2l$ dimensions for any $l$.

Our construction of these solutions exploits the connection
between non-commutative fields and operators in single particle 
quantum mechanics. Under this correspondence, 
the $\star$ product maps onto usual operator multiplication, and 
the equation of motion translates into algebraic operator equations.
The noncommutative scalar action can be rewritten as the trace over 
operators (which can be regarded as $\infty\times \infty$ matrices). This leads  
to a connection between noncommutative field theories, and zero dimensional 
matrix models.

Next we consider noncommutative $U(N)$
Yang-Mills theory. When expanded around a
simple solution of the equations of motion, 
the action takes the simple quartic form (up to constants and topological terms)
 \eqn\fjfl{S_{YM}={1\over 4 g_{YM}^2}\int d^{2l} x
\d^{\mu\lambda}\d^{\nu \rho} {\rm Tr} \left( [\Ph_\mu,
\Ph_{\nu}]
 [\Ph_{\lambda},\Ph_\rho ] \right),}where 
$\Ph_{\m}$ are $N \times N$ hermitian matrices and all 
commutators are constructed from the $\star$
product. Note that the kinetic term has been shifted away!
The usual space-time gauge symmetries act linearly as
unitary transformations on the fields $\Phi_\mu$, and the
$\Phi_\mu=0$ vacuum leaves even local gauge symmetries unbroken.
This construction is similar to that of \hlrs, in which the
kinetic term of Witten's string field theory action \wsft\ is
shifted away. Indeed, our search for such a construction in
noncommutative field theory was motivated by the tantalizing 
analogy, noted in \msv, 
between  noncommutative field theories and string field theories.
The existence of the formulation \fjfl\ of noncommutative 
gauge theories strengthens the analogy. 
We also reproduce, as an illustration, 
the $U(1)$ instanton solutions of \ns\ .

Rewriting noncommutative fields as the large $N$ limit of matrices, 
\fjfl\ is
closely related to the IKKT matrix theory
\refs{\ikkt}. Indeed, our construction is essentially equivalent to 
that presented by Aoki et. al.   
\aoki\ in this context.
\nref\rwl{N.Ishibashi, S. Iso, H. Kawai, Y. Kitazawa, 
``Wilson Loops in Noncommutative Yang Mills,''
hep-th/9910004.}%
\nref\rbm{I. Bars and  D. Minic, ``
Noncommutative geometry on a discrete periodic lattice and gauge theory, 
''hep-th/9910091. }%
\nref\rambjorna{J. Ambjorn , Y.Makeenko, J. Nishimura, R.Szabo, 
``Finite $N$ matrix models of noncommutative gauge theory, ''
hep-th/9911041,  JHEP { \bf 11} 029 (1999).}%
\nref\rambjornb{J. Ambjorn, Y. Makeenko, J. Nishimura, R. Szabo, 
``Nonperturbative dynamics of noncommutative gauge theory,''
  hep-th/0002158.}%
\nref\li{M. Li, ``Strings from IIB Matrices, '' hep-th/9612222,
Nucl.Phys. {\bf B499}, (1997), 149.}%
Related observations are also made in \refs{\li, \rcds, \rwl -\rambjornb}.

This paper is organized as follows. In section 2 we describe the
action for noncommutative scalar field theory. In section 3 we consider the
limit, $\t \to \infty$, in which the
equations simplify considerably. The general
solution can be found exactly and is given in terms of quantum
mechanical projection operators. In Section 4 we show that there are stable
solitons in this limit, as long as the potential has at least two local
minima. 
In section 5 we argue that there are stable solitons at large
but finite $\t$ which can be constructed perturbatively in $\t^{-1}$. In
section 6 we turn to the noncommutative gauge theory where the purely
quartic action is constructed. The $U(1)$ instanton solution of \ns\ 
is also reproduced. In an Appendix we give an explicit construction, 
of the leading ${1\over \t}$ correction to the 
simplest stable soliton of the scalar field theory.

\newsec{The Noncommutative Scalar Action}

Consider first a noncommutative field theory of a single scalar $\ph$
in $(2+1)$ dimensions with non-commutativity purely in the spatial directions. 
The spatial $R^2$ is parametrized
by complex coordinates $z, \bar{z}$. The energy functional
\eqn\act{E={1 \over g^2}\int d^2z \left( \p_{z}\ph\p_{\bar{z}}\ph + 
V(\ph) \right),}
where $d^2z=dx dy$. (We will comment on the generalization to arbitrary
dimensions in the appropriate places.)
Fields in this non-local action are multiplied using the Moyal star
product,
\eqn\stare{\left( A \star B \right)(z, \bar{z})=e^{{\t \over 2}
\left( \p_{z}\p_{\bar{z}^{\prime}}  -\p_{z^{\prime}}\p_{\bar{z}} \right)}
A(z, \bar{z})B(z^{\prime}, \bar{z}^{\prime})
|_{z_=z^{\prime}}.}
Note that in the quadratic part of the action, the star product reduces to
the usual product.

We seek finite energy (localized) solitons of
\act. These can also be interpreted as finite action
instantons in the two-dimensional euclidean theory.
We will, however, refer to the solutions
as solitons in the following.  

Since no solutions exist in the commutative limit $\t=0$ \der\ , we begin
our search in the limit of large noncommutativity,  $\t \r \infty$.
It is useful to non-dimensionalize the coordinates $z\r z\sqrt{\t}$,
$\bar{z}\r \bar{z}\sqrt{\t}$. As a result, the $\star$ product
will henceforth have no $\t$; i.e. it will be given by \stare\  with $\t=1$.
Written in rescaled coordinates, the dependence
on $\t$ in the energy is entirely in front of the potential term:
\eqn\actr{E={1\over g^2}\int d^2z \left( \half (\p\ph)^2 +\t V(\ph) \right)}
In the limit $\t \r \infty$, with
$V$ held fixed,
the kinetic term in \actr\ is negligible in
comparison to $V(\ph)$, at least for field configurations
varying over sizes of order one in our new coordinates.

Our considerations apply to generic potentials $V(\ph)$, but we will,
for definiteness, 
mostly discuss those of polynomial form
\eqn\potr{V(\ph)=\half m^2 \ph^2 + \sum_{j=3}^{r} {b_j \over j}
 \ph^j .} 
We have, of course,  abbreviated
$$\ph^j= \ph \star \ph\star \cdots \star \ph  .$$

\newsec{Scalar Solitons in the $\t=\infty$ Limit}

After neglecting the kinetic term, the energy 
\eqn\actrr{E= {\t \over g^2}\int d^2z V(\ph),}
is extremised by solving
the equation
\eqn\inft{{\p V \over \p \ph}=0. } For instance,
\inft\ is
\eqn\cub{m^2\ph+b_3 \ph\star\phi=0}
for a cubic potential and
\eqn\quar{m^2\ph+b_3\ph\star\phi+b_4\phi\star\phi\star\phi=0}
for a quartic potential.

If $V(\ph)$ were the potential in a commutative scalar field theory,
the only solutions to \inft\ would be the constant configurations
\eqn\trivsoln{\ph= \l_i ,}
where
$\l_i\in \{\l_1 , \l_2, \cdots ,\l_k\}$ are the various real extrema of the
function $V(x)$\foot{
For $V(\ph)$ as in \potr, $\l_{i}$ are
the real roots of the equation
${m^2 x  + \sum_{j=3}^{r} b_j x^{j-1}=0. }$}.
As we shall see below, the derivatives in the definition of the
star product allow for more
interesting solutions of \inft.

\subsec{ A Simple Nontrivial Solution}

A non-trivial solution to \inft\
can easily be constructed. Given a function $\ph_0(x)$ that obeys
\eqn\bas{ \left( \ph_0 \star \ph_0 \right)(x)=\ph_0(x),}
it follows by iteration that $\ph_0^n(x)=\ph_0(x)$,\foot{This equation and
its solution has
also appeared in earlier work involving the Moyal Product. See \refs{\moyalb
,\moyalc }.}   
and that
$f\left( a \ph_0(x) \right)= f(a) \ph_0(x)$ (fields in
$f$ are multiplied using the star product).
In particular, $\l_i \ph_0(x)$
solves  \inft\ when $\l_i$ is an extremum of $V(x)$.
Thus, in order to find a solution  of
\inft, it is sufficient to find a function that squares to itself under
the star product. We proceed to construct such a function below.

If we take the ordinary product of a smooth function of width $\Delta$ with
itself, the spatial size of the function shrinks to a fraction of $\D$,
which is why non-constant functions never square to themselves!
The non-locality of the star product, however,
introduces an additional effect, adding roughly\foot{ 
The added width is actually $\approx K$, 
the typical  momentum in the Fourier transform of the function.
For a function of size $\D$ with no oscillations,  $K \approx {1 \over \D}$.
For a function of size $\D$ with $n$ oscillations, $K \approx {n \over
\D}$.}
${1 \over \D}$
to the width of the
product.
This makes it possible
for a lump of approximately unit size to
square to itself under the star product.

Consider a gaussian packet of the form
$$\psi_{\D}(r)= {1\over \pi \D^2} e^{-{r^2 \over \D^2}},$$
with radial width $\D$ (here $r^2=x^2+y^2$).
The star product of $\psi_{\D}$ with itself is easily computed
by passing to momentum space,
\eqn\ft{\tilde{\psi}_{\D}(k)= \int e^{ik\cdot x}
\psi_{\D}(x) d^2x=e^{-{k^2 \D^2 \over 4}},}
\eqn\fts{\eqalign{\left(\tilde{\psi}_{\D} \star \tilde{\psi}_{\D} \right)(p)&=
{1 \over (2 \pi )^2 }
\int d^2k \tilde{\psi}
_{\D}(k)\tilde{\psi}_{\D}(p-k)e^{{i \over 2} \ep_{\m\n}k^{\m}
(p-k)^{\n}}\cr &={1 \over 2\pi \D^2}e^{-{p^2 \over 8}  \left(
{\D^2}+ {1 \over \D^2} \right)}. }}
Therefore
\eqn\sng{\left( \psi_{\D} \star \psi_{\D}\right)(x)=
{ 1\over \pi^2 \D^2 (\D^2+ {1 \over \D^2} )}
\exp\left[{ -{2r^2 } \over
{\D^2}+ {1 \over \D^2} }\right].}
In particular\foot{
We note in passing that in the limit $\D \r 0$, \sng\ reduces to
$\d^2(x) \star \d^2(x)= {1 \over (2 \pi)^2}$.},
 when $\D^2=1$, the gaussian squares to itself (up to a factor of $2\pi$).
That is,
\eqn\soln{ \ph_0(x)=2\pi \psi_{1}(x)=
2e^{-r^2}}
solves \bas\ and $\l_i \ph_0(x)$ solves
\inft.

\subsec{The General Solution}

%\cl{{\it  The Strategy}}

In order to find all solutions of \inft\ we will exploit the
connection between Moyal products and quantization.
Given a  $C^\infty$
function $f(q,p)$ on $R^2$ (thought of as the phase space of a one-dimensional
particle), there is a prescription which
uniquely assigns to it an operator $O_f({\hat q},{\hat p})$,
acting on the corresponding single particle
quantum mechanical Hilbert space, $\cal{H}$.
It is convenient for our purposes to choose
the Weyl or symmetric ordering prescription
\eqn\weyl{ O_f({\hat q},{\hat p}) = { 1\over (2\pi)^2}\int d^2k
\tilde{f}(k) e^{-i\left( k_q {\hat q} +k_p {\hat p} \right)},}
where
\eqn\ft{\tilde{f}(k)=\int d^2x e^{i(k_q q +k_p p)} f(q,p),}
and
\eqn\comrel{[{\hat q},{\hat p}]=i.}
With this prescription,  it may be verified
%\foot{This may be
%shown as follows. The RHS is $\int {d^2z\over
%2\pi}\langle z|O_f(a^{\dagger},a)|z\rangle$ where $|z \rangle$ is a
%coherent state. But $\int {d^2z\over
%2\pi}\langle z|O_f(a^{\dagger},a)|z\rangle=\int
%{d^2z\over2\pi}f_N(z,\bar{z})
%={\tilde{f}_N(0) \over 2\pi} = {\tilde{f}(0) \over 2 \pi}=
%{1\over 2 \pi} \int d^2z f(z,\bar{z})$, where
%$f$ is the function associated with $O$ by \weyl\ and $f_N$ is the
%%normal ordered symbol associated with $O$, defined above eq ??}
that
\eqn\vol{{1\over 2\pi}\int dpdq f(q,p)=\Tr_{\CH}O_f ,}
and that the Moyal product
of functions is isomorphic to ordinary operator multiplication
\eqn\opst{O_{f}\cdot O_{g} = O_{f\star g}.}

In order to solve any algebraic equation involving the star product,
it is thus sufficient to determine all operator solutions to the
equation in $\CH$.
The functions on phase space corresponding to each of these operators may then
be read off from \weyl. We will now employ this procedure to find
all solutions of \inft.

As noted above, any solution to \bas\ may be rescaled into
a solution of \inft. Particular solutions of \inft\ may thus be obtained
by constructing operators in $\CH$ that obey \bas, i.e.
$O_{\ph}^2=O_{\ph}$. This equation is solved by any projection operator in
$\CH$. $\CH$ possesses an infinite number of projection operators, which
can be classified by the dimension of the subspace they project onto.
Each class contains a large continuous infinity of operators,
each of which, upon rescaling, yields a solution to \inft.

The most general solution to  \inft\ hence takes the form
\eqn\solnarb{ O=\sum_{j}a_jP_j}
where $\{P_{j} \}$ are mutually orthogonal projection operators onto one
dimensional subspaces, with
$a_j$ taking values in the set $\{\l_i\}$ of real extrema of $V(x)$.

In order to obtain the functions in space corresponding to the solutions
\solnarb, it is convenient to choose a particular basis in
$\CH$. Let $|n\rangle$  represent the energy eigenstates of
the one dimensional harmonic oscillator whose creation and annihilation
operators are defined by
\eqn\aad{a={{\hat q}+i{\hat p} \over \sqrt{2}};
\  \ a^{\dagger}={{\hat q}-i{\hat p}  \over \sqrt{2}}.}
Note that
$a|n\rangle =\sqrt{n}|n-1\rangle$  and
$ a^{\dagger}|n\rangle=\sqrt{n+1}|n+1\rangle$.
Any operator may be written as
a linear combination of the basis operators $|m\rangle \langle n|$'s,
which, in turn, may be expressed in terms of $a$ and $a^{\dagger}$ as
\eqn\asp{|m\rangle\langle n|=:{a^{\dagger m} \over \sqrt m!} e^{-a^{\dagger}a}
{a^{n} \over \sqrt{n!}}:}
where double dots denote normal ordering.

We will first describe operators of the form \solnarb\ that correspond to
radially symmetric functions in space. As $a^{\dagger} a \approx {r^2 \over
2}$, operators corresponding to radially symmetric wavefunctions are
functions of $a^{\dagger} a$. From \asp, the only such operators are
linear combinations of
the diagonal projection operators $|n\rangle \langle n|
={1 \over n!}:a^{\dagger n}e^{-a^{\dagger}a}a^n: $.
Hence all radially symmetric solutions of \inft\
correspond to operators of the form
$O=\sum_n a_n |n\rangle \langle n|$, where the numbers $a_n$ can take
any values in the set $\{ \l_i\}$.

We now translate these operator solutions back to field space. From 
the Baker-Campbell-Hausdorff formula
\eqn\norma{  e^{-i\left( k_q {\hat q} +k_p {\hat p} \right)}=
e^{-i\left( k_{\bar{z}} a +k_z a^{\dagger} \right)}
=e^{-{k^2 \over 4}}:e^{-i\left( k_{\bar{z}} a +k_z a^{\dagger} \right)}: ,}
where
$$k_{z}={ k_x +ik_y \over \sqrt{2}}, \  \ k_{\bar{z}}=
{ k_x -ik_y \over \sqrt{2}}, \  \ k^2=2k_z k_{\bar{z}}. $$
Any operator $O$ expressed as a normal ordered function of $a$ and
$a^{\dagger}$, $f_{N}(a, a^{\dagger})$,
can be rewritten in Weyl ordered form as follows.
By definition,
\eqn\nw{O=:f_N(a, a^{\dagger}):=
{ 1\over (2\pi)^2}\int d^2k
\tilde{f}_N(k) :e^{-i\left( k_{\bar{z}} a +k_z a^{\dagger} \right)}: .}
Using \norma, \nw\ may be rewritten as
\eqn\nwt{O=
{ 1\over (2\pi)^2}\int d^2k
\tilde{f}_N(k)e^{{k^2 \over 4}}
e^{-i\left( k_{\bar{z}} a +k_z a^{\dagger} \right)}.}
Thus, the momentum space function
$\tilde{f}$ associated with the operator $O$, according to
the rule \weyl\ is
\eqn\nwt{\tilde{f}(k )=e^{{k^2 \over 4}} \tilde{f}_N(k).}
For the operator $O_n =|n\rangle \langle n|$ we find, using \asp\ and
\nw, that the corresponding normal ordered function 
$\tilde{\ph}^{(n)}_N(k)=2\pi e^{{-k^2 \over 2}}L_n({k^2 \over 2})$.
\nwt\ then becomes
\eqn\nnm{|n\rangle \langle n|= { 1\over (2\pi)}\int d^2k
e^{{-k^2 \over 4}}L_n({k^2 \over 2})
e^{-i\left( k_{\bar{z}} a +k_z a^{\dagger} \right)}}
where $L_n(x)$ is the $n^{th}$ Laguerre polynomial.
The field $\ph_n(x,y)$ that corresponds to the operator
$O_n = |n\rangle \langle n|$ is, therefore,
\eqn\phn{\ph_n(r^2=x^2+y^2)= { 1\over (2\pi)}\int d^2k
e^{{-k^2 \over 4}}L_n({k^2 \over 2})
e^{-ik.x}=2(-1)^ne^{-r^2} L_n(2r^2).}
Note that $\ph_0(r^2)$ is precisely the gaussian solution found in Sec. 3.1.

In summary, \inft\
has an infinite number of real radial solutions, given by
\eqn\sumsol{\sum_{n=0}^{\infty} a_n \ph_n(r^2)}
where $\ph_n(r^2)$ is given by \phn\ and each $a_n$ takes
values in $\{\l_i\}$.

In order to generate all non radially symmetric solutions to
\inft,  we rewrite \actrr\ in operator language, using \vol\ as
\eqn\actopr{E= {2 \pi \t \over g^2}\Tr V(O_{\ph}).}
\actopr\ is manifestly invariant under unitary transformations of
$O_{\ph}$ and so has a $U(\infty)$ global symmetry. In other words, if $O$
is a solution to the equation of motion, 
so is $U O U^{\dagger}$, where $U$ is any unitary
operator acting on ${\cal H}$.
A general Hermitian
operator (one that corresponds to a real field $\ph$)
may be obtained by acting on a diagonal operator (i.e. an
operator that corresponds to a radially symmetric field configuration)
by an element of the $U(\infty)$ symmetry group (since any hermitian
operator is unitarily diagonalizable). Thus every solution to \inft\
may be obtained from a radially symmetric solution
by means of $U(\infty)$ symmetry transformations.

Therefore solutions to \inft\ consist of disjoint infinite dimensional
manifolds labelled by the set of eigenvalues of the corresponding operator.
Points on the same manifold can be mapped into each
other by $U(\infty)$ transformations.
Each manifold includes several\foot{Distinct diagonal operators
having the same eigenvalues lie on the same manifold, being related by the
``Weyl'' subgroup of $U(\infty)$ that permutes eigenvalues.}
diagonal operators (radially symmetric
solutions). We will have more to say about the moduli space of these
solutions in the next section.

As all solutions are related to radially symmetric solutions
by a symmetry transformation, we will mostly discuss only radially
symmetric solutions.

\subsec{UV/IR Mixing}

$\ph_0(r^2)$, the Gaussian solution worked out in subsec 3.1,
is a lump of unit size centred at the origin, as
shown in Fig. 1.
\fig{A plot of $\ph_0(r)$ versus $r$. The solution is a blob
centred at the origin.}{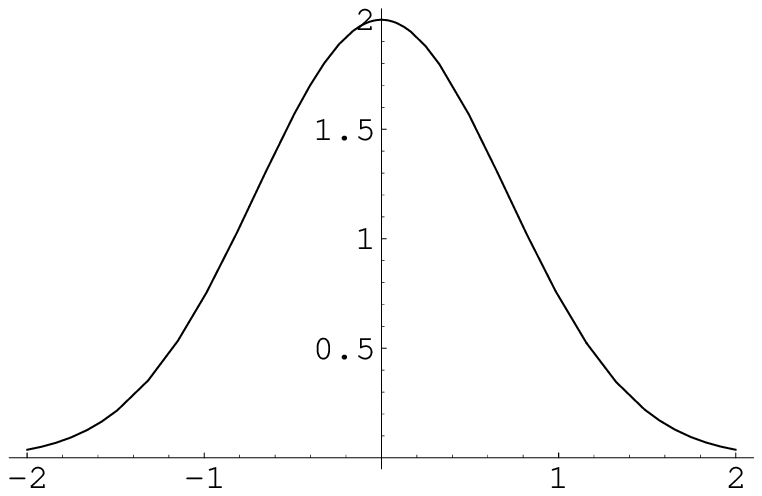}{2.0truein}
$\ph_n(r)$,  at large $n$, looks quite different (see Fig. 2.).
It is a solution of size $\approx \sqrt{n}$ that undergoes
$n$ oscillations\foot{
Using asymptotic formulae for Laguerre polynomials we find
$$\ph_n(r)= \cases{
2(-1)^n& $ r \ll \sqrt{{1\over 8n}}$ \cr
{2 (-1)^n \over ( 2 \pi^2 r^2)^{{1\over 4}}}\cos(\sqrt{2n}2r-{\pi\over 4})
& $\sqrt{{1\over 8n}} \ll r \ll \sqrt{2n})$\cr
{2(-2r^2)^n \over n!} e^{-r^2} & $ r \gg \sqrt{2n}$\cr.}$$}
in that interval, with
oscillation period $\propto {1 \over \sqrt{n}}$.
$\ph_n(r^2)$ thus receives significant
contributions from momenta up to $\sqrt{n}$ in momentum space. These
solutions  exemplify the UV-IR mixing pointed out in \msv; oscillations
with frequency $\sqrt{n}$ produce an object of size $\sqrt{n}$
(instead of ${1\over \sqrt{n}}$) in a noncommutative theory.
\fig{A plot of $\ph_{30}(r)$ versus $r$.}{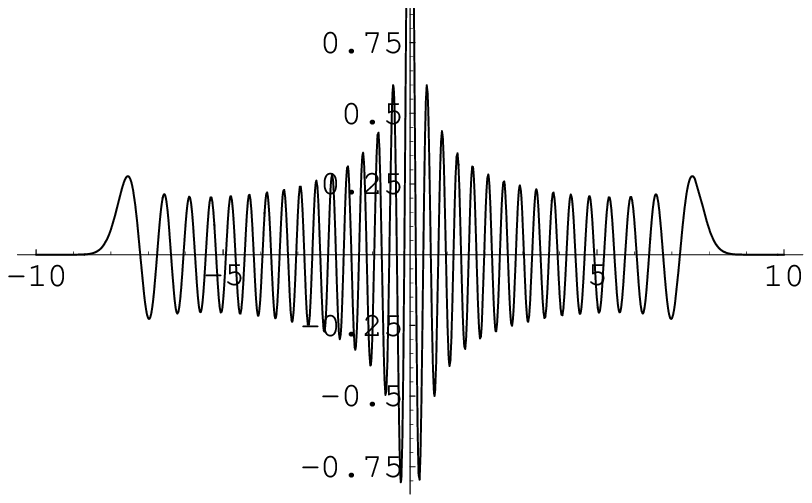}{2.5truein}

\subsec{Generalization to Higher Dimensions}

All considerations of the preceding subsections may 
easily be generalized to higher dimensions. 
Consider a scalar field theory in $2l+1$ dimensions with non-commutativity
only in the spatial directions. 
By a choice of axes, the $2l \times 2l$ dimensional noncommutativity 
matrix $\T$ may always be brought into block diagonal form.    
In other words,  it is possible to choose 
spatial coordinates $z_i, {\bar z}_{\bar
j}$ $(i,{\bar j}=1...l)$, in terms of which the non-commutativity matrix 
$\T_{i{\bar j}}=\t_i\d_{i{\bar j}}$, $\T_{i j}=\T_{\bar{i},\bar{j}}=0$.,
As before we consider the limit where $\t_i$ are uniformly
taken to $\infty$ and non-dimensionalize 
$z_i\r z_i\sqrt{\t_i}$.
As in the previous subsections, the kinetic term in the action may be 
dropped in this limit. Solutions to the equations of motion \inft\ are
once again in correspondence with operator solutions to the same 
equations; the operators in question now acting on 
 $\CH \times \CH \times...\times \CH$, $l$ copies of the Hilbert space of the 
previous subsection. 
The general solution to \inft\ once again takes the 
form \solnarb\ in terms of projection operators on this space.
As in the previous subsection, the general solution may be obtained from 
diagonal solutions via $U(\infty)$ rotations.  
Diagonal solutions to \inft\ are given by 
\eqn\arbdim{O = \sum_{\vec{n}}a_{\vec{n}}|\vec{n}\rangle\langle \vec{n}|
\leftrightarrow \sum_{\vec{n}}a_{\vec{n}}\prod_{i}\ph_{n_i}(|z_i|^2),}
where $\vec{n}$ is shorthand for the set
 of quantum numbers $\{n_i\}$ 
for the $l$ dimensional oscillator and $\phi_{n_i}$ are 
defined in \phn. As in
\sumsol, the coefficients $a_{\vec{n}}$ take values in $\{\l_i\}$.
A subset of the solutions \arbdim\ are actually invariant under $SO(2l)$  
rotations 
and can be written in terms of associated Laguerre polynomials. These are
displayed in Sec.A.3 of the Appendix.

In summary, in the limit of maximal noncommutativity, the construction 
of solitons in two spatial dimensions generalizes almost trivially to 
every even spatial dimension. 

\newsec{Stability and Moduli Space at $\t=\infty$} 

In this section we study the stability of the solitons constructed in 
the previous section. We will also describe the moduli space of stable
solitons. 

\subsec{Stability at $\t=\infty$}

We wish to examine the stability of the radial solution
\eqn\radwoln{ \ph(r^2)=\sum_{n=0}^{\infty}\l_{a_n} \ph_n(r^2)}
to small fluctuations. 
Since any $U(\infty)$ rotation does not change the energy of our solution 
\radwoln,
it is sufficient
to study the stability of \radwoln\ to radially symmetric fluctuations.
These are most conveniently parameterized as deformations of the
eigenvalues.
The energy for an arbitrary radially symmetric state
$ \ph(r^2)=\sum_{n=0}^{\infty}c_n\ph_n(r^2)$ is
$$E={2\pi \t \over g^2}\sum_{n=0}^{\infty} V(c_n).$$
The solution
$c_n=\l_{a_n}$ is manifestly an extremum of $S$, as, by definition,
$\l_{a_i}$ are extrema of the function $V(x)$.
Clearly \radwoln\ is a local minimum of the energy
(and so a stable solution) if, and only if,
$\l_{a_n}$ is a local minimum of $V(x)$ for all $0 \leq n \leq \infty$.

As an example consider the cubic potential of Fig. 3. with a maximum at
 $\l=-1$. In this case, all $\l_{a_n}$ in \radwoln\ are either zero or
-1. The only stable solution is that for which all $\l_{a_n}=0$, i.e.
the vacuum.
The  solution $-\ph_0(r^2)$, for instance, is unstable, as the
energy of this field configuration is decreased by scaling this
solution by a constant near unity. This instability shows up as
a negative eigenvalue of the quadratic form for fluctuations about
$-\ph_0(x)$; the corresponding eigenmode
$\d\ph_0$ is $ \propto \ph_0$.
\fig{The $\ph^3$ theory with an unstable extremum}{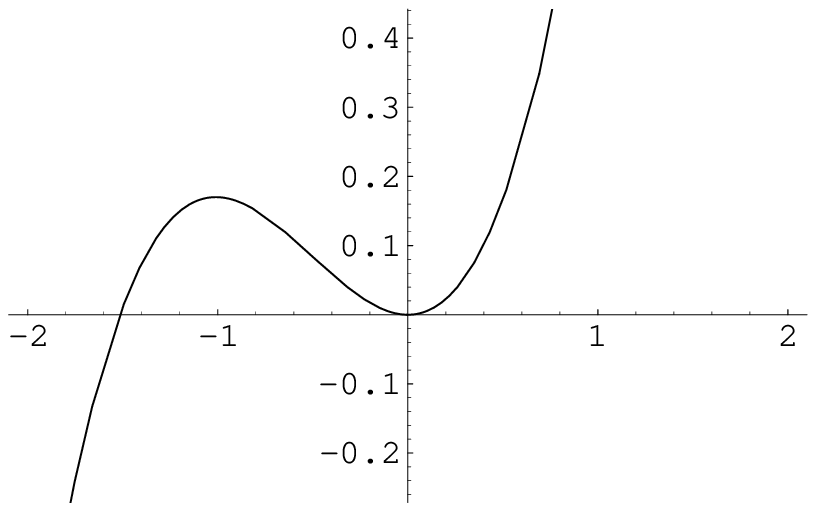}{2.0truein}

On the other hand the field theory with $V(\ph)$ (say, for a quartic
potential) graphed in
Fig. 4 has stable solitons; these are solutions of the
form $\ph(r^2)= \sum_{n=0}^{\infty} \l_{c_n} \ph_n(r^2)$
with $\l_{c_n}$ taking the values of the minima -- $0$ or $\l \approx -1.4$
for all
$n$.
In particular
$\l \ph_0(r^2)$ is a stable solution, manifestly stable to rescalings.
Again, one may check that the quadratic form for fluctuations
about $\l \ph_0(r^2)$ is positive. In particular, $\d\ph \propto \ph_0$
is an eigenmode of this quadratic form with positive eigenvalue.
\fig{A $\ph^4$
potential with two minima.}{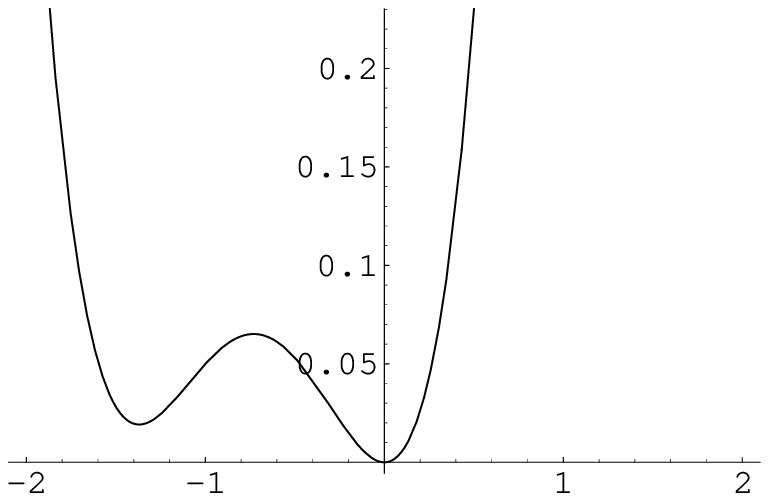}{2.0truein}
The stability of $\ph(r^2)= \l \ph_0(r^2)$ in the previous example may
qualitatively be understood as follows.
$\ph_0$ is a Gaussian of height $2\l$. Far away from the origin,
$\ph_0(x)=0$, but near $x=0$,  $\ph_0(x)$ is in the vicinity of the
second vacuum.
In other words, the static
solution corresponds to a bubble of the ``false'' vacuum. The area of the
bubble is of order one (or $\t$ in our original coordinates),
the non-commutativity scale.
In a commutative theory such a bubble would decay by shrinking to zero
size. Noncommutativity prevents the bubble from shrinking to
a spatial size smaller than
$\sqrt{\t}$. In order to decay, $\ph_0$ actually has to scale to zero - but
that process involves going over the hump in the potential
and so is
classically forbidden.
\fig{Profile of the
Gaussian soliton with a false vacuum region (above the horizontal bar)
of radius 1.}
{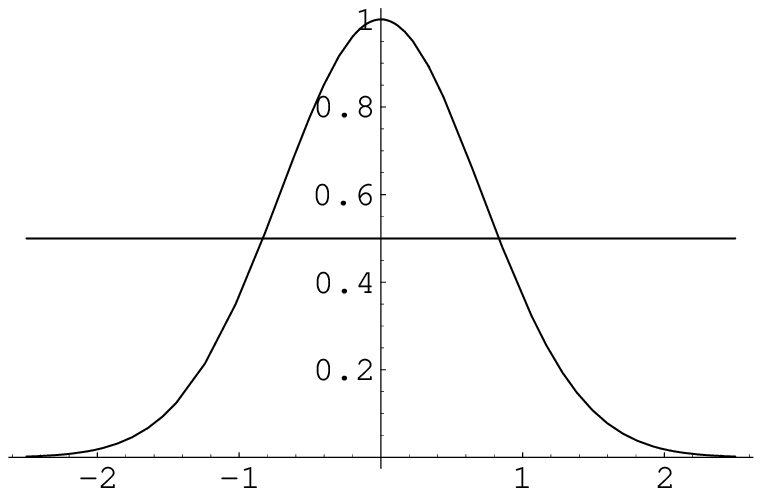}{2.0truein}
The energy of this soliton is proportional to 
the vacuum energy  density ${V(\l)\over g^2}$ at the `false' vacuum 
times the volume  of the soliton $\t$. 
It is remarkable that   
the energy of the soliton is completely insensitive to the value of the 
scalar potential at any point except $\ph=\l$. Thus the mass
of the soliton is unchanged if the height of the 
barrier in $V(\ph)$ (between $\ph=\l$ and $\ph=0$, 
see Fig. 4.) is taken to infinity while $V(\l)$ is kept fixed.  
This is true even though $\ph_0(r)$, the solitonic field configuration
corresponding to $\l |0\rangle \langle 0|$, decreases continuously from 
$\ph=2 \l$ at $r=0$ to $\ph=0$ at $r= \infty$!
  
Consider a 2+1 dimensional scalar theory, noncommutative only 
in space, at infinite  $\t$. Using the correspondence between 
functions and operators (matrices) described in the previous section, 
the noncommutative 
scalar field theory is equivalent to the matrix quantum 
mechanics of an $N \times N$ hermitian matrix $H$, at infinite $N$, 
with the usual relativistic kinetic term $\Tr \left( \p_t H \right)^2$, 
and a potential $\Tr\left( V(H) \right)$. The amplitude for an  
eigenvalue of $H$ to tunnel from $\l$ to $0$ is exponentially suppressed by the 
area under the potential barrier in Fig. 4., and goes to zero as this 
area is taken to infinity. Thus the finite mass 
soliton $\l |0\rangle \langle 0|$
is stable, even quantum mechanically,  in this limit.

 The $U(\infty)$ symmetry of \actrr\ is spontaneously broken
by every nonzero
solution, $\ph(x)$, of  \inft. As a consequence, every solution
has a number of exact
zero modes (Goldstone modes)
corresponding to
small displacements about $\ph(x)$ on the manifold of solutions.
As $R_{nm}=|n\rangle \langle m|+|m \rangle \langle n|$ and
$S_{nm}=i(|n\rangle \langle m|-|m \rangle \langle n|)$ are the generators
of $U(\infty)$,  these zero modes are given by the nonzero elements of
$\d\ph \propto [R_{nm}, \ph], [S_{nm}, \ph]$.

The $U(\infty)$ group of symmetry transformations
that generates these zero modes is certainly not manifest (at least
to the untrained eye) in the energy written in coordinate
space in the form \actrr. In addition to the two translations,
\actrr\ possesses three manifest local
symmetries, corresponding to a linear change of the coordinates
$x,y$ by an SL(2,R) matrix. The remaining $U(\infty)$
transformations act non-locally on
$\ph(x,y)$, according to $\ph'(x,y)=\left( U \star \ph \star U^{\dagger}
\right) (x,y)$ where $U(x,y)$ is any function that obeys 
$U \star U^{\dagger}=1$ (such functions correspond to $U(\infty)$ 
operators under the map \weyl).

All arguments in this subsection may be applied (after straightforward 
generalizations) to higher dimensional solitons.

\subsec{Multi Solitons}

In this subsection we will qualitatively describe
a part of the moduli space of stable
solitons (at $\t=\infty$)
in the simple case of the potential graphed in Fig. 4 with a single non-zero
minimum at $\ph=\l$.

The stable solitons can be characterized by their `level'
(number of $\l$ eigenvalues). All stable level one solitons correspond
to operators of the form
\eqn\lon{\l U |0 \rangle \langle 0|U^{\dagger}}
where $U$ is a  unitary operator. As mentioned above, the set of
level one solitons span an infinite dimensional manifold
parameterized by $U(N) / U(N-1)$ (for $N=\infty$).

The soliton looks very different at different points on the manifold.
$U=I$ in \lon\ corresponds to  the gaussian blob of Fig. 1.
If $U$ happens to be a unitary transformation that
maps $|0 \rangle $  to $|m \rangle$, for large $m$, the corresponding
wave function is qualitatively similar to that in Fig. 2. 
When $U=e^{a^{\dagger}z- a\bar{z}}$ is the generator of translations,
the operator in \lon, $\l |z \rangle \langle z|$, is proportional
to the projection operator onto a gaussian  centred around
$z={1\over \sqrt{2}}(x+iy)$. (Here $|z \rangle=e^{-{|z|^2\over 2}}
e^{a^{\dagger}z}|0 \rangle$ is the usual coherent state.)
Again, if $U$ corresponds
to one of the $SL(2,R)$ operators, we obtain squeezed states;
gaussians elongated in the
$y$ direction and shrunk in the $x$ direction. And so on.

Turn now to solitons at arbitrary level $n$. All such solitons may be
obtained by acting on
$$ \l( \ph_0+\ph_1 + \cdots + \ph_{n-1})$$
by arbitrary unitary transformations.
The manifold of solutions thus generated is parameterized by
$ {U(N) \over U(N-n)} $
(and has dimension $d_n \approx 2nN$) where $N \r \infty$.

Notice that $d_n \approx n d_1$. This fact has a nice explanation;
in a particular limit the manifold of level $n$ solutions reduces
to $n$ copies of level $1$ solitons very far from each other.
This conclusion follows from the observation that the operator
that represents $n$ widely separated level one solitons (with centres
$z_j$), for instance
\eqn\approxmulti{M=\l \sum_{j} |z_j\rangle \langle z_j|}
is approximately a level $n$ soliton (and exponentially
close to a true level $n$ soliton) when $|z_i-z_j| \r \infty$
for all $i, j$. We demonstrate this explicitly below for the case
$n=2$.

Using $\langle z|-z\rangle=e^{-2|z|^2}$, it is easy to check that
the kets
\eqn\tort{|z_{\pm} \rangle ={|z \rangle \pm |-z \rangle \over \sqrt{2(1 \pm
e^{-2|z|^2})}}}
are orthogonal. From \solnarb\ we conclude that the projector
\eqn\lts{O_z=\l \left( |z_+\rangle \langle z_+ |+ |z_-\rangle \langle z_-|
\right)=\l {|z \rangle \langle z | +|-z\rangle \langle-z|
+  e^{-2|z|^2}\left(|z \rangle \langle-z | +|-z\rangle \langle z| \right)
\over {(1-e^{-4|z|^2})} }}
corresponds to
a level 2 solution. Up to corrections of order $e^{-2|z|^2}$,
$O_z$ is equal to $|z \rangle \langle z | +|-z\rangle \langle-z|$,
the superposition of field configurations
corresponding to two widely separated level one solitons
\foot{It is curious that the kinetic energy of this field configuration is
independent of $z$ indicating that there is no force between the two
solitons even to next leading order in ${1\over \theta}$. }. 
We conclude that a part of the level $n$ moduli space describes
$n$ widely separated level one solitons.

We have, so far, worked in the strict limit $\t=\infty$.
The picture developed in this limit is
 qualitatively modified at large but finite $\t$, as we will describe
in the next section.

\newsec{Scalar Solitons at Large but Finite $\t$.}

We have argued that, under certain conditions on $V(\ph)$,
\inft\ has an infinite number of stable solutions. Each
solution has an infinite number of exact zero modes,
the Goldstone modes of the spontaneously broken
$U(\infty)$ symmetry of \actrr.

At finite $\t$, the
 kinetic term in \actr\ explicitly breaks this $U(\infty)$ symmetry
down to the Euclidean group in 2 dimensions.
Finite $\t$ effects may thus be expected to
\item{1.} Lift the $\t=\infty$ manifold of solutions to
a discrete set of solutions.
\item{2.} Give (positive or negative) masses to the $U(\infty)$
Goldstone bosons about these discrete solutions.

In Appendix A we will argue that, 
at large enough $\t$, corresponding to every radially
symmetric solution $s$ of \inft, there is a radially symmetric saddle point of
\actr, that reduces to $s$ as $\t \r \infty$. It is likely that 
these are the only saddle points of \actr. 

Not all these radially symmetric solutions are stable, however.
In fact, it might seem likely that some of the infinite number of zero modes,
at $\t=\infty$, about each solution $s$, might become tachyonic at finite $\t$. 
If this were true,
\actr\ would have no classically stable extremum at any finite $\t$,
no matter how large.

We will find that is not the case. In subsection 5.1 below we will argue
that any small perturbation of \actrr\ must preserve the existence of
at least one classically stable level one soliton. In subsection
5.2 we will identify this soliton to be the one near the gaussian 
$\l\ph_0(r^2)$.

\subsec{Existence of A Stable Soliton}

For definiteness, through the rest of this section we assume
that the potential
$V(\ph)$ has the shape shown in Fig. 4. In particular, it is positive
definite. 
Let the stable extremum of
$V$ occur at $\ph=\l$ and the unstable extremum at $\ph=\b$,
($\l< \b < 0$).

Consider, first, \actrr\ i.e. the energy functional in the limit where we neglect
the kinetic term.
We will show that any path in field space leading from
the soliton $ \l \ph_0(r^2)$ to
the the vacuum passes through a point whose energy
is larger than ${2 \pi \t\over g^2} V(\b)$. Since the energy of the stable soliton
is ${2 \pi \t\over g^2} V(\l)<{2\pi\t \over g^2} V(\b)$, 
every path from
the soliton to the vacuum must 
pass over a barrier of height $\CO({\t\over g^2} )$.

The energy evaluated on an operator $A$ is
\eqn\fnv{E= {2\pi\t \over g^2} \Tr(V(A))={2\pi\t \over g^2} \sum_{n=1}^{\infty} V(c_n)}
where $c_n$ are the
eigenvalues of $A$. Since $V$ is positive
definite,
\eqn\bbgsa{E \geq {2\pi\t \over g^2}  V(b) ,}
where $b$ is the smallest eigenvalue of $ A$.

Consider a path in field, or operator space, leading from
$\l \ph_0$ to the vacuum. At the beginning of this path
$b=\l$. At its end $b=0$. Since $\l<\b<0$, any smooth path
with these endpoints must have a
point at which $b=\b$. At that point 
$E>{2\pi\t \over g^2} V(\b)$, as was to be shown.

Now include the the kinetic term in \actrr.
Barring singular behaviour, this changes the energies of all field
configurations by terms of $\CO({1 \over g^2})$.
For large enough $\t$, the arguments of the previous paragraph
imply that the field configuration that describe the level one
soliton at $\t=\infty$ cannot decay to the vacuum. Hence there must exist
at least one stable soliton near one of the unperturbed
level one solutions. In fact, as we will show in the next subsection,
there is a stable soliton near the gaussian $\l\ph_0(r^2)$. In the Appendix 
we will present an approximate construction of this solution at large but 
finite $\t$. A similar argument demonstrates the existence of at least one
stable solution at level $n$.

\subsec{ Approximate Description of the Stable Soliton}

All level one solutions to \inft\ take the form
$\l U|0 \rangle \langle 0|U^{\dagger}$ where $U$ is a unitary operator.
We wish to determine the contribution of the kinetic term to the
energy of such an operator.

The kinetic term in \act\  for an  operator $A$ is
\eqn\kinetic{K={2 \pi \over g^2} \Tr [a, A][A, a^{\dagger}].}
Setting $A=\l U  |0 \rangle \langle 0| U^{\dagger}$ we find
\eqn\kinrot{ { g^2 K(U) \over 2 \pi \l^2}=1+\sum_k 2k|U_{k,0}|^2 -2|\sum_k
\sqrt{k+1}U_{k,0}U^*_{k+1,0}|^2 .}

We expand \kinrot\ to quadratic
order in deviations from $U=I$. Choose
$U_i=U_{i,0}$ for $i \geq 1$ as the coordinates for this expansion
($|U_{00}|$ is determined in terms of $U_i$ as $U$ is unitary).
To quadratic order in $U_i$
\eqn\kinrotq{ { g^2 K(U) \over 2 \pi \l^2}=1+2\sum_{k=2}^{\infty} k |U_{k}|^2.}
As $U_1$ and $\bar{U}_1$ do not appear in \kinrotq, they
parameterize flat directions of $K(U)$ (to quadratic order).
This was to be expected.
Any localized extremum of \act\ has two exact translational zero
modes. Infinitesimally, $U_{01}$ and its complex conjugates act as
derivatives on $\ph_0(r^2)$, generating these zero modes.
Modulo these zero modes, the
fluctuation matrix about $U=1$ is positive definite.

While $K(U)$ has several critical points
other than $U=I$, it has no further local minima. For example,
$U=U^{(m)}$, the unitary transformation that rotates
$|0 \rangle \langle 0|$ to  $| m \rangle \langle m|$, is an
unstable critical point of $K(U)$ for all $m$. In fact $U=U^{(m)}$
is unstable to decay into $U=I$. 
This
may be demonstrated by considering the path in field space
$|\a\rangle\langle\a|$ where $|\a \rangle = \cos{\a}|0\rangle+\sin{\a}|m\rangle$.
\kinetic\ evaluated on such a path is equal to $1+2m\sin^2{\a}$ (for $m>1$;
$1+2\sin^4{\a}$ for $m=1$)
indicating that the state $| m \rangle \langle m|$ can decay to 
$|0 \rangle \langle 0|$. 

We will now argue that, at large enough $\t$,
 the finite $\t$ saddle point $\ph(x,y)$ of
\actr\ that reduces to $\l |0\rangle \langle 0|$ as $\t \r \infty$
is classically stable.

Consider the mass matrix for fluctuations about $\ph(x,y)$.
Since any operator may be
written as $U D U^{\dagger}$ where $D$ is diagonal and $U$ unitary,
small fluctuations may be decomposed into radial (fluctuations
of $D$) and angular ones (fluctuations of $U$). The mass matrix for purely
radial fluctuations is $\CO(\t)$ to leading order, and has been
shown to be positive definite in sec 3.4. The mass matrix for
purely angular fluctuations is $\CO(1)$ to leading order,
and has been shown to be positive definite, modulo the two zero modes.
Since angular modes completely disappear from the potential, mixing
between radial and angular fluctuations occurs only through the
kinetic term, and are also $\CO(1)$. These cross terms result in
corrections to the eigenvalues of the mass matrix only
at $\CO({1 \over \t})$.
Hence, to leading order in ${1 \over \t}$, the mass matrix
is positive. The two zero
modes of the angular mass matrix cannot be driven negative
by ${1 \over \t}$ corrections as they are exact.

A similar argument demonstrates the instability of all other radially
symmetric level one
solitons (those that reduce to $\l |n\rangle \langle n|$ at $\t=\infty$)
at large enough $\t$.

The considerations of this subsection may easily be generalized 
to solitons in $2l$ spatial dimensions, 
using the higher dimensional analogue of \kinrot : 
\eqn\gennn{{ g^2K(U) \over (2 \pi)^l \l^2}
=1+2\sum_{j,\vec{k}}k_j|U_{\vec{k},\vec{0}}|^2
-2\sum_j|\sum_{\vec{k}}
\sqrt{k_j+1}U_{\vec{k},\vec{0}}U^*_{\vec{k}+\vec{i},\vec{0}}|^2 }
and \kinrotq\
\eqn\gemnnn{{ g^2K(U) \over (2 \pi)^l \l^2}=
1+2\sum_{\vec{k}}\left(\sum_{j=1}^{l}k_j-\sum_{j=1}^{l}\d_{\vec{k},\vec{i}} \right)
|U_{\vec{k},\vec{0}}|^2 .}
We use the notation of \arbdim; $\vec{k}$ is an $l$ dimensional 
vector, $j$ runs from $1$ to $l$ and 
$\vec{i}$  is the basis unit vector in the $i^{th}$ direction; in  
components $i_n= 
\d_{i,n}$.
Notice that $K(U)$ in \gemnnn\ is independent of $U_{\vec{i},0}$
for all $i$, a consequence of the exact translational invariance in 
all $2l$ spatial directions.

\newsec{Noncommutative Yang-Mills}

\subsec{Quartic Action for the $U(1)$ Theory in Two Dimensions}

Consider the action
\eqn\fdl{S= {1\over 4g_{YM}^2}\int d^2 z  [\bar \Phi,  \Phi][\bar \Phi,\Phi],}
where $\Phi$ is a complex field and
\eqn\jga{ [\Phi, \bar \Phi] \equiv  \Phi\star \bar \Phi-\bar \Phi\star \Phi.}
The equation of motion following from \fdl\ is
\eqn\dosc{ [\bar \Phi , [\Phi, \bar \Phi] ]=0.}
$\Phi$ can also be viewed as a quantum mechanical operator
and $\bar \Phi$ as it's hermitian conjugate. The commutators
in \fdl-\dosc\ are then ordinary operator commutators, and the 
integral is
the trace over the Hilbert space.
In the operator representation a
simple solution of the equation of motion \dosc\ is
\eqn\phs{\Phi=a,~~~~~\bar \Phi =a^{\dagger}.}
Let us expand around this solution by defining
\eqn\phes{\Phi=a+iA_{\bar z},~~~~~\bar \Phi =a^{\dagger}-iA_{z}.}
One then finds, translating back to functions (with
$\sqrt{2} z=q+ip$, $[a,~]=\p_{\bar z}$ and $[a^\dagger,~]=-\p_z$, that
\eqn\trax{[\Phi, \bar \Phi]=1+ i\p_zA_{\bar z}-i\p_{\bar z}A_z -[A_z,A_{\bar
z}]= 1+ iF_{z\bar z}.}
The operator representation of \fdl\ has the manifest 
$U(N=\infty)$ symmetry under which 
$\Phi \r \Phi^{\prime}= U^{\dagger}\Phi U$ just as
in the scalar field theory. 
Infinitesimally,
\eqn\gaugesym{\d\Phi=i[\Phi,\L],}
where $U=\exp{i\L}$. When gauged, this is just 
the usual $U(1)$ gauge symmetry of the non-commutative theory, 
$$\d A = d\L +i[A,\L].$$

The equation of motion \dosc\ is
\eqn\eoms{D_{ \bar z}F_{z \bar z}=0.}
The action \fdl\ is then
\eqn\fhl{S =-{1\over 4g_{YM}^2}\int d^2 z \bigl(F_{z \bar z}-i\bigr)^2,}
the standard two dimensional non-commutative $U(1)$
Yang-Mills action up to constants and topological terms.

\subsec{The U(N) Theory in $2l$ Dimensions}

\fdl\ can be generalized to \eqn\fjl{S ={1\over 4g_{YM}^2}\int d^{2l} x
\d^{\mu\lambda} \d^{\nu \rho} {\rm Tr} \bigl( [\Phi_\mu,
\Phi_{\nu}] [\Phi_{\lambda},\Phi_\rho ] \bigr) ,} where
$\mu,\nu=1,,,2l$ and $\Phi_\mu $ are real $N\times N$ matrices.
Though we have restricted ourselves to a flat euclidean metric, one can
generalise the argument below to the Minkowski metric as well.

The equation of motion is \eqn\fazx{\d^{\mu\nu }  [
\Phi_{\mu}, [\Phi_{\nu},\Phi_\lambda ]]=0.} We choose complex coordinates
such that $\Theta_{a \bar b}=i\d_{a \bar b},$ 
with $a,b=1...l$. \fazx\ has the solution
\eqn\fty{\Phi_b=a_b, ~~~~~~~~~\Phi_{\bar b}=a^\dagger_{\bar b},}
where $[a_b,a^\dagger_{\bar c}]=\d_{b \bar c}.$
Expanding around this solution with 
\eqn\fty{\Phi_b=a_b+iA_{\bar b}} one finds 
\eqn\fzdl{S =-{1\over 4g_{YM}^2}\int
d^{2l} z \bigl( F_{a{\bar b}} -\Theta^{-1}_{a{\bar b}} \bigr)^2.}
As before the manifest 
$U(\infty)\otimes U(N)$ 
symmetry corresponds to the non-commutative $U(N)$ gauge symmetry.

\subsec{The U(1) Instanton}

The four dimensional non-commutative gauge 
theory has instanton solutions which are deformed
versions of the usual non-abelian instantons. In particular, the $U(1)$
non-commutative theory also has non-singular finite action saddle points
\ns. We exhibit the operators $\Phi_a$ corresponding to the simplest such 
$U(1)$ instanton.

The operators $\Phi_a$ corresponding to an anti self dual field strength 
$\d^{a{\bar b}}F_{a{\bar
b}}=0$ $(a,{\bar b}=1,2)$, obey
\eqn\sd{[\Phi_b,\Phi_c]=0,~~~~~~~~
\d^{a \bar b}[\Phi_{a},\Phi_{\bar b}]=2.}

In four dimensions, the operators $\Phi_a (a=1,2)$ live in a
Hilbert space generated by the creation and annihilation operators of a
two-dimensional
harmonic oscillator (See Sec. 3.3). 
Rather than work in the conventional number basis
$|n_1, n_2\rangle$, it is convenient to work in Schwinger's angular momentum
basis,
\eqn\jmbasis{|j,m\rangle\equiv {(a^{\dagger}_1)^{j+m}\over \sqrt{(j+m)!}}
{(a^{\dagger}_2)^{j-m}\over \sqrt{(j-m)!}}|0,0\rangle  ,}
with $0\leq j <\infty , |m|\leq j$.
The operators
\eqn\angmom{J_{+}=a^{\dagger}_1a_2,~~~ J_{-}=a^{\dagger}_2a_1 ,~~~ J_{z}=
\half (a^{\dagger}_1a_1 - a^{\dagger}_2a_2)}
obey the usual angular momentum algebra.

We will find a solution to \sd\ of the form 
\eqn\ansatz{\eqalign{\Phi_b&= a_b\sum_{j,m}(1+c_j)|j,m\rangle\langle j,m|
=a_b +a_b\sum_{j,m}c_j|j,m\rangle\langle j,m|,\cr
\Phi_{\bar b}&= a^{\dagger}_{\bar b},}}
and put it into Hermitian form via a complexified gauge transformation $W$. 

The ansatz \ansatz\ satisfies
the holomorphic part of \sd\ for any $c_j$. For a real
$c_j$, the only condition comes from the equation
$F_{1{\bar 1}}=-F_{2{\bar 2}}$. Using
\eqn\aadag{\eqalign{a^{\dagger}_{1,2}|j,m\rangle &= \sqrt{j\pm m+1}|j+\half, m\pm \half
\rangle ;\cr  a_{1,2}|j,m\rangle &= \sqrt{j\pm m}|j-\half, m\mp \half \rangle ,}}
yields the equation $jc_j=(j+1)c_{j+\half}$. Which has the solution
\eqn\slf{c_j={c\over j(2j+1)},~~~~(j>0).}
The complexified gauge transformation 
\eqn\yis{W=W^{\dagger}= \sum_{j,m}\sqrt{j\over j+1}|j,m\rangle }
puts the solution \ansatz\ into Hermitian form for $c=-1$. 
The field strength then takes the compact form
\eqn\fmn{[\Phi_{\bar b},\Phi_{c}]=-\d_{\bar b c}-iF_{\bar b c}=
-\d_{\bar b c}-(\vec{J}\cdot\vec{\sigma})_{\bar b c}
\sum_{j,m}{1\over j(j+1)(2j+1)}|j,m\rangle\langle j,m|.}
Here $\vec{J}$ are the angular momentum generators defined in \angmom\ and
$\vec{\sigma}$, the usual Pauli matrices. This solution is exactly the same as
the simplest charge one $U(1)$ instanton in \ns.
It may be checked that ${\half}TrF_{a {\bar b} }^2 = 1$.

\bigskip

\centerline{\bf Acknowledgements} We are grateful to S. Coleman,
J. Harvey, T. Kinoshita, J. Maldacena, B. Pioline, N. Seiberg, I. Singer, C.
Vafa, M. Van Raamsdonk, S. Sinha, A. Vishwanath
 and S.-T. Yau for useful discussions. This work was supported
in part by DOE grant DE-FG02-91ER40654.

\appendix{A}{ Solutions at  Finite $\t$}

In this appendix we will examine radially symmetric saddle points 
of \actr\ at finite $\t$. In subsection A.1 we study the equation of 
motion resulting from \actr\ at finite $\t$, and examine the existence 
of radially symmetric solutions to these equations. In A.2 we concentrate
on a particular solution; the one that reduces to the stable soliton 
$\l |0\rangle \langle 0|$ as $\t$ is taken to infinity. We present an 
approximate construction of this soliton at large $\t$. In 
A.3 we briefly comment on the generalization of these results 
to solitons in higher dimensions.   

\subsec{The Perturbation Expansion and a Recursion Relation}

The full equation of motion derived from \act\ may be written 
in momentum space as
\eqn\fl{ \tilde{\ph}(k^2) +\sum_{j=3}^{r}{b_j \over m^2}\tilde{\ph}^{j-1}(k^2)=
{-k^2 \over m^2\t} \tilde{\ph}(k^2)}
While the LHS of \fl\ is independent of $\t$, the RHS is of order
${1 \over \t}$, and so is a small parameter at large $\t$. For
notational convenience,
we set ${b_j \over m^2}=d_j$ and ${1\over m^2 \t}=\ep$.

Let
\eqn\sugsol{\sum_{n=0}^{\infty} c_n \tilde{\ph}_n(k^2)}
be a solution to \fl. Substituting \sugsol\ into \fl,
using the recurrence relation for Laguerre polynomials, and equating
coefficients of $\tilde{\ph}_n(k^2)$,  we arrive at the difference equations
\eqn\difference{ c_n+\sum_{j=3}^{r}d_jc_n^{j-1}=2\ep [ n c_{n-1} -(2n+1)c_n
+(n+1)c_{n+1}].}
We are interested in finite energy solutions to \inft, i.e. solutions to
\difference\ for which
\eqn\cond{  \sum_{n} V(c_n)<\infty . }
Since $V(0)=0$, \cond\ will be satisfied if the $c_n $s approach
zero sufficiently fast as $n$ approaches infinity. For such a solution,
all nonlinear terms in \difference\ may be neglected at
large enough $n$. At sufficiently large $n$, $n$ may also be replaced
by a continuous variable $u$, and \difference\ turns into the second
order differential equation
\eqn\didi{c(u)=2\ep u {d^2c(u) \over du^2}.}
\didi\ is the Schroedinger equation for a zero energy state of a particle
in a ${1\over u}$ potential. $\sqrt{\ep}$ plays the role of Planck's constant,
and at small $\ep$ \didi\ is easily solved in the WKB approximation, yielding
\eqn\didisoln{c(u)={ A_{-}u^{{1\over 4}}e^{-\sqrt{2u \over \ep}} }
+
A_{+}u^{{1\over 4}}{e^{+\sqrt{2u \over \ep}} }
}
where $A_{\pm}$ are arbitrary constants. In order that
$c_n$ tend to zero at large $n$, $A_{+}=0$.
Thus, for large\foot{ \didisoln\ is a good approximation when $|c_n| \ll 1$
(so that dropping nonlinear terms in \difference\ is justified)
and ${c_n-c_{n-1}\over
c_n} \ll 1$, i.e. $n \ep \gg 1$ (so that the transition from \difference\
to   \didi\ is justified).} $n$,
\eqn\assol{c_n \approx  An^{{1\over
4}}{ e^{-\sqrt{{2n \over \ep}} }} .}
\assol\ has an undetermined parameter $A$, the scale of the solution
at large $n$. As \difference\ is a nonlinear equation, $A$ is not an
arbitrary parameter, but is determined to be one of a discrete set of
values. Given $c_p$ and $c_{p+1}$, the $(p+1)$ equations 
\difference\ with $n=0\cdots p$ overdetermine the $p$ unknowns $c_n$ for $n<p$. 
The extra 
equation constrains the scale $A$, as we'll see in the next subsection.  
%
%\eqn\differencez{ c_0+\sum_{j=3}^{r}d_j c_0^{j-1}=2\ep [-c_0
%+c_1],}
%
%Choose some $p \gg {1 \over \ep}$. We wish to determine all solutions to
%\difference\ that are well approximated by the form \assol\ for
%all $n\geq p$, i.e. solutions for which $|c_n| \ll 1$ for $n \geq p$.
%At $\ep=0$ all such solutions are given by the $s^p$ functions of
%the form \sumsol, (where $s$ is the number of real extrema of $V(x)$.).
%These are simply all solutions which have $c_n=0$ for $n>p$.
%At nonzero $\ep$, solutions are more complicated, but may 
%be determined, as we will see in the next subsection.

%Consider any set $\{c_i \}$  which give a
%solution to \fl\ of the form \sugsol\  and satisfy the
%conditions of the previous paragraph.
%By hypothesis, $c_p$ and $c_{p+1}$ are specified  by \assol, in terms
%of the unknown parameter $A$.
%Using \difference\ for $n=p$, $ c_{p-1}$ may be determined as a
%function of $A$. Continuing this procedure recursively, $c_n$ for
%all $n< p$ may be determined; this procedure terminates when
%$c_0$ is determined by \difference\ with $n=1$. This leaves
%\difference, with $n=0$
%as a constraint equation. This is a polynomial
%of degree $(r-1)^p$ in $A$ and may be used to solve for $A$.
%Each real root of
%\differencez\ corresponds to a real solution to \difference.
%These are the solutions that the $s^p$ solutions of the
%$\ep=0$ equation flow to at finite $\ep$.

\subsec{The Gaussian Soliton Corrected}

In this section we present an approximate construction of the 
stable soliton that reduces to the gaussian at infinite $\t$. 
Our construction approximates the 
true solution to arbitrary accuracy at small enough $\ep$.

We wish to find a solution of \difference\ such that
\eqn\cone{\lim_{\ep \r 0} c_0=\l }
and
\eqn\ctwo{\lim_{\ep \r 0}c_m =0}
uniformly in $m$, for $m\geq 1$. \ctwo\ ensures that, on such a solution,   
\difference\ for $n \geq 1$   reduces to 
\eqn\differencel{ c_n=2\ep [ n c_{n-1} -(2n+1)c_n
+(n+1)c_{n+1}]}
for small enough $\ep$. It is easy to find an explicit solution to 
\differencel\ that obeys \cone, \ctwo. Consider a function $\ph(x,y)$ that
obeys the differential equation
\eqn\difcons{(-\ep \p^2 +1) \ph=b \ph_0.}
Expanding $\ph$ in the form 
\eqn\sugsol{\ph=\sum_{n=0}^{\infty} c_n \ph_n}
and imitating the manipulations of section 4.3, we find that 
$c_n s $ obey \differencel\ for $n \geq 1 $, but obey
\eqn\dz{ c_0 =2 \ep \left[ c_1-c_0 \right]+ b}
instead of \difference\ (with $n=0$). This relation 
will fix the free parameter 
$b$. 

\difcons\ is easily solved in momentum space 
\eqn\sodic{\tilde{\ph}(k)=b {\tilde{\ph}_0(k) \over1+ 2\ep k^2}. } 
Using the explicit
forms for $\tilde{\ph}_n(k)$ and orthogonality of the Laguerre polynomials we find
\eqn\sonlnd{ c_n=b \int_{0}^{\infty} dx {e^{-x} L_n(x) \over 1+2 \ep x}.}
In particular                             
\eqn\dzs{c_0=b \int dx{ e^{-x} \over 1+2\ep x}=
b F(\ep) \ {\rm where} \ F(\ep)=1-\ep + \CO(\ep^2).}
Using \dzs\ we conclude that \dz\ and \difference\ (at $n=0$) are identical
on $\{ c_n \}$ if $b$ is chosen such that  
\eqn\condid{b F(\ep)+\sum_{j=3}^{r}d_j
\left(b F(\ep) \right)^{j-1}=b\left(F(\ep)-1 \right).}
We wish to find a solution to \condid\ that 
obeys \cone, i.e. (from \dzs) one for which $\lim_{\ep \r 0}b= \l$. 
As $\l +\sum_{j=3}^{r}d_j\l^{j-1}=0$, such a solution exists, 
and takes the form
\eqn\ss{b(\ep)=\l(1 + K \ep + \CO(\ep^2))} 
at small $\ep$ where $K$ is a number that may easily be determined. 

In summary, $\{ c_n \}$ given by \sonlnd\ with $b$ given by \condid, \ss, 
solve \difcons\ for $n \geq 1$ and \difference\ (with $n=0$). 
$\{ c_n \}$ therefore also approximately satisfy the true difference equations
\difference\ for all $n$  as long as $|c_n| \ll 1$ for all $n \geq 1$. 
But it is easy to verify that all $|c_n|$ for all $n \geq 1$ are 
arbitrarily small at small enough $\ep$.
Using the completeness of the Laguerre polynomials, 
\eqn\sqsum{ \sum_{n=0}^{\infty} c_n^2=b^2 \int_{0}^{\infty} {e^{-x} \over 
(1+2\ep x)^2} < b^2.}
But 
\eqn\dzb{c_0=b \int_{0}^{\infty} dx {e^{-x} \over 1+2\ep x}
> b \int_{0}^{\infty} dx e^{-x}(1-2\ep x)=b(1-2\ep).}
Combining \sqsum\ and \dzb\ 
\eqn\allsmall{ \sum_{n=1}^{\infty} c_n^2 <4\ep b^2}
establishing \ctwo\ uniformly in $n$ on our solution.
Thus $\{ c_n \}$ provides an approximate solution to 
the full nonlinear difference equations \difference\ for 
all $n$ at small enough $\ep$. Furthermore, from \sqsum, 
this solution has finite energy. 

As $\{c_n\}$ obey the linearized recursion relation \difcons\ and are
small at small $\ep$, we can conclude, from the previous subsection,
that $d_n$ 
takes the form \assol\ for $n\ep \gg 1$. In order to estimate the 
behaviour of $c_n(\ep)$ for $n \ll {1 \over \ep}$ we formally expand the 
denominator in \sonlnd\  in a power series in $\ep x$ and integrate
term by term, arriving at the asymptotic expansion
\eqn\asdn{c_n=\sum_{m=n}^{\infty} (-1)^{m+n} (2 \ep)^m {m!^2 \over 
n! (m-n)!}.} 
This expansion is useful only when the first few terms 
in the series in \asdn\ are successfully smaller, i.e. for $n\ep \ll 1$. 

\subsec{Generalization to Higher Dimensions}

In this subsection we will outline the generalization 
of the arguments of A.1 and the construction of A.2, 
for the case of the maximally isotropic 
noncommutativity in $2l$ dimensions, i.e. a theory with noncommutativity
matrix $\T$, all of whose eigenvalues are $\pm i \t$. 
It is likely that these arguments can be further extended to generic
$\T$.

We first note that a subset of the diagonal
$\t=\infty$ solutions \arbdim\ 
are 
(in non-dimensionalized coordinates) invariant under $SO(2l)$  
rotations. These solutions take the form
\eqn\radlag{\sum_{\vec{n}}{c_J}
{1 \over \sqrt{D_J}}\d_{\left(J,\sum_{i} n_i \right)}
|\vec{n}\rangle\langle \vec{n}|
\leftrightarrow {1 \over \sqrt{D_J}}\sum_{J}{c_J }\ph_J^{(l)}(r^2).}
Here 
\eqn\spfor{\ph_J^{(l)}(r^2=\sum_i|z_i|^2)=2^l(-1)^JL_J^{(l-1)}(r^2).}
where $L_J^{(l-1)}(r^2)$ is an  associated Laguerre polynomial.
\spfor\ is obtained from \arbdim\ by repeated use of the identity
$$\sum_{m=0}^{n}L_{n-m}^{\a}(x)L_{m}^{\b}(y)=L_n^{\a+\b+1}(x+y).$$
$D_J=\pmatrix{J+l-1 \cr J}$ is a convenient normalization factor. 

When the noncommutativity matrix is maximally isotropic, the 
kinetic term in \actr\ is invariant under $SO(2l)$ rotations of 
rescaled coordinates. Thus the corrections to an $SO(2l)$ invariant 
$\t=\infty$ solution, of the form \radlag,  are also 
$SO(2l)$ invariant. 

Restricting to $SO(2l)$ invariant functions, the arguments of 
section A.1 are easily generalized. Any $SO(2l)$ invariant function 
takes the form
 \eqn\momexp{\tilde{\ph}(k^2)=\sum_{n=0}^{\infty}c_J\tilde{\ph}_J(k^2);\qquad
 \tilde{\ph}_J(k^2)={1\over
 \sqrt{D_J}}(2\pi)^lL_J^{(l-1)}({k^2\over2})e^{-{k^2\over 4}}.}
The equation of motion implies that $c_J$ obey the following generalization 
of \difference\ 
 \eqn\diffarb{c_J+\sum_{j=3}^{r}d_jc_J^{j-1}=2\ep [ (J+l-1) c_{J-1} -(2J+l)c_J
 +(J+1)c_{J+1}].}
For large $J$ \diffarb\ and \difference\ are identical, hence all 
conclusions of section A.1 carry over to this case.

The perturbative construction of the solution that reduces 
to the $SO(2l)$ invariant Gaussian proceeds as in section A.2 yielding the 
approximate result (good for small $\ep$)
\eqn\dnarb{d_J={b \over \sqrt{D_J}\Gamma(l)}\int_{0}^{\infty} dx 
 {x^{l-1}e^{-x} L_J^{(l-1)}(x) \over 1+2 \ep x}.}

\listrefs
\end